\documentclass[twoside]{article}

\usepackage{lipsum} % Package to generate dummy text throughout this template

\usepackage[sc]{mathpazo} % Use the Palatino font
\usepackage[T1]{fontenc} % Use 8-bit encoding that has 256 glyphs
\usepackage{microtype} % Slightly tweak font spacing for aesthetics

\usepackage[hmarginratio=1:1,top=32mm,columnsep=20pt]{geometry} % Document margins
\usepackage{multicol} % Used for the two-column layout of the document
\usepackage[hang, small,labelfont=bf,up,textfont=it,up]{caption} % Custom captions under/above floats in tables or figures
\usepackage{booktabs} % Horizontal rules in tables
\usepackage{float} % Required for tables and figures in the multi-column environment - they need to be placed in specific locations with the [H] (e.g. \begin{table}[H])
\usepackage{hyperref} % For hyperlinks in the PDF

\usepackage{lettrine} % The lettrine is the first enlarged letter at the beginning of the text
\usepackage{paralist} % Used for the compactitem environment which makes bullet points with less space between them
\usepackage{graphicx}
\graphicspath{ {/} }
\usepackage{abstract} % Allows abstract customization
\usepackage{titlesec} % Allows customization of titles
\renewcommand\thesection{\Roman{section}} % Roman numerals for the sections
\renewcommand\thesubsection{\Roman{subsection}} % Roman numerals for subsections
\titleformat{\section}[block]{\large\scshape\centering}{\thesection.}{1em}{} % Change the look of the section titles
\titleformat{\subsection}[block]{\large}{\thesubsection.}{1em}{} % Change the look of the section titles

\usepackage{fancyhdr} % Headers and footers
\title{\vspace{-15mm}\fontsize{24pt}{10pt}\selectfont\textbf{Vulnerability Analysis of GWireless}} % Article title

\author{
\large
\textsc{Benjamin Lim (A0100223)}\\[2mm] % Your name
\normalsize National University of Singapore \\ % Your institution
\normalsize limbenjamin@u.nus.edu \\ % Your email address
}
\date{April 21, 2015}

\begin{document}

\maketitle % Insert title

\thispagestyle{fancy} % All pages have headers and footers

%----------------------------------------------------------------------------------------
%	ABSTRACT
%----------------------------------------------------------------------------------------

\begin{abstract}

\noindent Wireless networking has become very popular in recent years due to the increase in adoption of mobile devices. As more and more employees demand for Wi-Fi access for their devices, more companies have been jumping onto the "Bring Your Own Device" (BYOD) bandwagon\cite{1} to appease their employees. One such example of an enterprise wireless infrastructure is the George Washington University's GWireless.
\\
\\For this project, I will attempt to capture hashes of authentication credentials from users who are connecting to the GWireless network using what is commonly known as the "evil twin" attack. I will document the hardware, software used and steps taken to configure the devices. I will then evaluate the feasibility of such an attack, explore variations of the attack and document measures that can be taken to prevent such an attack.

\end{abstract}

%----------------------------------------------------------------------------------------
%	ARTICLE CONTENTS
%----------------------------------------------------------------------------------------

\begin{multicols}{2} % Two-column layout throughout the main article text

\section{Introduction}

\lettrine[nindent=0em,lines=2]{M}any organizations worldwide turn to WPA-Enterprise (802.1x) standard when implementing a secure wireless network. Some of the merits of 802.1x include
~\\
\begin{compactitem}
\item Auditability - Each user logs on using his own credentials instead of a common password when using WPA-PSK mode.
\item Interoperability - All major operating systems support 802.1x without the need for additional software.
\item Authentication - All variants of 802.1x include Extensible Authentication Protocol (EAP). As part of the EAP handshake process, the client would verify the certificate sent by the server, hence ensuring its identity.
\end{compactitem}
~\\

GWireless uses 802.1x standard as well. However, as far as I know, GWU does not release the server certificates. As shown in the screenshot below, the instructions for connecting to GWireless also do not mention the need to configure certificates of any type. Therefore, there is no way for client devices to verify the authenticity of an access point.

\includegraphics[width=150px,keepaspectratio]{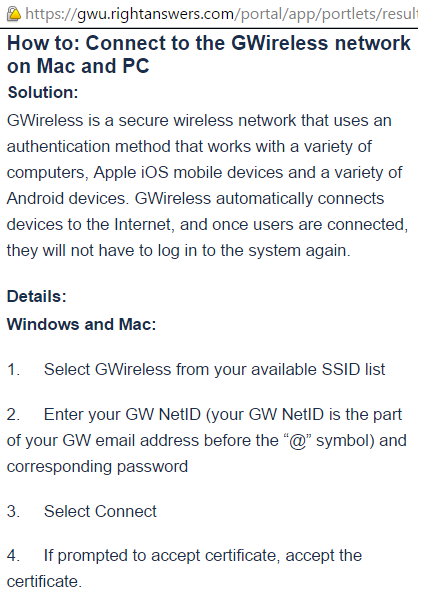}

%------------------------------------------------
\pagebreak

\section{Background}

The 802.1x standard has support for many combinations of EAP configurations. Some of the more common configurations are Extensible Authentication Protocol - Transport Layer Security (EAP-TLS) and Protected Extensible Authentication Protocol - Microsoft Challenge Handshake Authentication Protocol v2 (PEAP-MS-CHAPv2). EAP-TLS is widely recognised to be the more secure among the two\cite{2}. It supports mutual authentication in which both client and server are issued a cert and both parties verify the identity of the other party before performing further authentication. However, it is more rarely seen as compared to PEAP-MS-CHAPv2 which is a password based system. For PEAP-MS-CHAPv2, the client verifies the identity of the server by ensuring that the the certificate chain leads to a root CA which is trusted by the machine. The server authenticates the client through the use of a challenge response mechanism to prove that the client knows the username/password combination. However, specifying server certificate is optional, if not configured on a client, the client will trust any access point (AP) that broadcasts the same service set identifier (SSID) and uses the same configuration.  

GWireless uses the PEAP-MS-CHAPv2 configuration and does not provide the server certificate for download\cite{3}. Thus clients connecting to GWireless will have no way to verify the identity of an access point.

%------------------------------------------------

\section{Method}

To perform this attack, I have relied on off-the-shelf consumer equipment that is readily accessible. The equipment consists of a single laptop running Ubuntu 14.04 as well as an Asus RT-N15U router, an entry grade wireless router. As pictured below, the case was stripped out for a previous project, but no hardware or software modification was done to the router.

\includegraphics[width=180px,keepaspectratio]{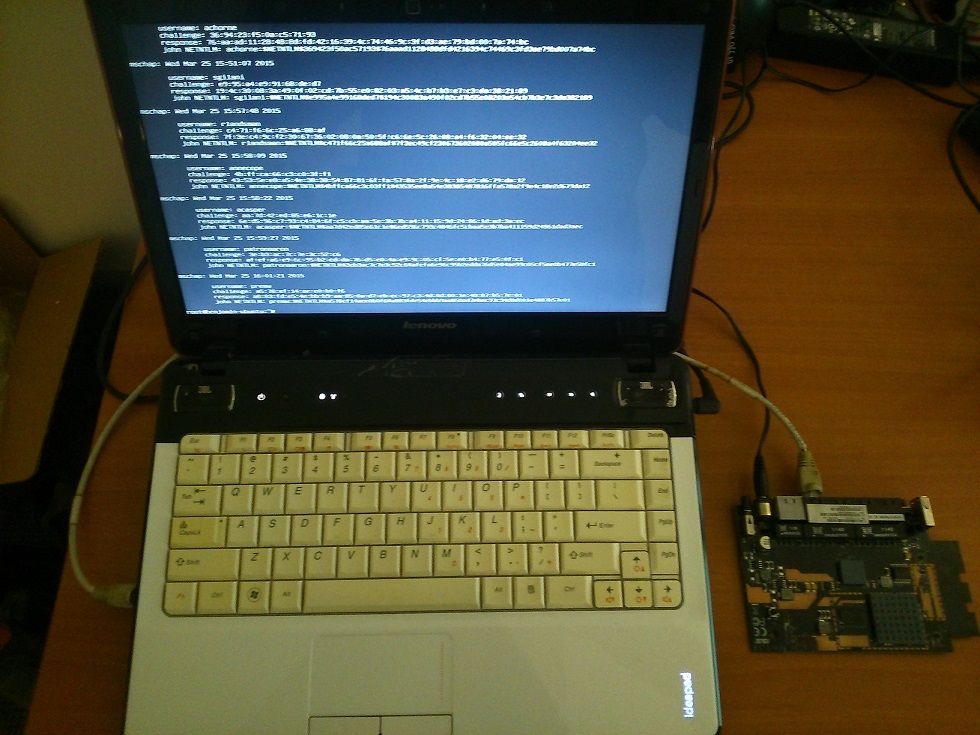}

For the RADIUS server, I have used a modified version of FreeRadius\cite{4} that has been patched to log down challenge and response pairs from client authentication attempts. FreeRadius is run as a daemon on the laptop and listens to a specified port. The laptop was configured to use a static IP address.

The router is connected to the laptop using an ethernet cable. The router's default firmware supports RADIUS authentication. It was set to AP mode and the radius server's IP was set to the laptop's IP. The shared secret was set to the same value as the config file on the laptop. The SSID was changed to "GWireless" and the security mechanism was set to PEAP/MS-CHAPv2.

Once the setup was done, any wireless device in the vicinity that has previously been associated with "GWireless" will try to connect to the rogue AP if the signal from the rogue AP is stronger than the real AP. This was one of the reasons why I chose to use an actual router instead of a soft AP like hostapd. The wireless signal emitted from a soft AP would likely be too weak for the attack to work. The connection to the rogue AP will fail because the radius server does not have the plaintext password and thus cannot complete the challenge response. As a result, the client device would then try to connect to another AP broadcasting the same SSID. Thus, this attack will likely remain undetected. The only possible giveaway is that it takes a slightly longer time to connect to the network. 

%------------------------------------------------

\section{Results}

I set up the equipment as specified at the ground floor of the Science and Engineering Hall (SEH) and proceeded to successfully collect 25 sets of credentials over a period of 90 minutes. Included in each set were the username, MS-CHAPv2 hash and the challenge response pair. The credentials can be found together with this report.

%------------------------------------------------

\section{Feasibility}

Up until this point, obtaining the hash required minimal amount of resources. An adversary only requires a laptop and a router that supports 802.1x which can be purchased for under \$500. All software used is free and open source. However, breaking the hash would require slightly more resources as described below.

\subsection{Non-targeted attacks}

An adversary can place the access point at a well trafficked location to gather as many hashes as possible and perform and offline dictionary attack later. Cassola et al. \cite{5}, performed a similar experiment on 17 CS graduate students and managed to crack the first password after 30 seconds and the 2nd after 2 hours using a 24 Xeon-CPU server. Considering the ease of obtaining a much larger sample size as well as the fact that the average user is likely to have a less secure password than a CS graduate student, it is highly likely that a desktop machine would suffice to crack a single hash in a reasonable amount of time (<24 hours).

\subsection{Targeted attacks}

The modus operandi for a targeted attack would be slightly different. In this case, the adversary would be interested in capturing the hash of a user with elevated privileges, hence he would try to get as close as possible to the target while running the fake AP to get the target's devices to authenticate with it. Once the target's handshake has been captured, he can proceed to perform an offline attack. Pico Computing\cite{6} built an FPGA box that is capable of breaking DES encryption within 24 hours. They have also offered a cloud service to crack all MS-CHAPv2 handshakes with a success rate of 100\% for only \$20. Hence such an attack would be highly feasible as well.
~\\
\\In conclusion, both non-targeted and target attacks are feasible and can be carried out by an adversary on a low budget.

%------------------------------------------------

\section{Attack Impact}

The GW NetID is being used for multiple purposes ranging from authenticating to GWireless, accessing GW email accounts and the Blackboard. An adversary who has obtained a user's credentials would be able to gain access to all these services. If the user is a student, an adversary will be able to view his assignment grades and submit assignments on his behalf, thus compromising his privacy. Obtaining the credentials to a professor's account could possibly allow a student to change his/her grade, thus presenting a larger problem. If the credentials belong to a systems administrator, the adversary would likely have even greater privileges.

An adversary would also be able to access all emails in the account and impersonate the user. He could also access or post illegal content on the internet which would implicate the user when law enforcement investigates.

Apart from that, it is likely that the staff members are able to access restricted file servers and printers using their NetID credentials and these resources would similarly fall into the adversary's hands.

%------------------------------------------------
\section{Variations of attack}

\subsection{Deauthentication attack}

A deauthentication attack involves sending a packet to dissociate clients from a particular basic service set identifier (BSSID). The client will then automatically try to reconnect to the GWireless network. If the rogue AP's signal is the strongest, the client will then attempt to connect to the rogue AP.  A deauth attack is especially useful if client device's roaming aggressiveness is low. Roaming aggressiveness is a measure of how often a client device will check and try to connect to an access point with stronger signal.

Therefore, a deauth attack will allow an adversary to capture more handshakes in a shorter period of time. The deauth attack should be executed while the rogue AP is running. For ease, the laptop running the radius server can also be used to mount a deauth attack. An adversary would have to install the open source aircrack-ng\cite{7} suite and use the aireplay-ng command to launch a deauth attack. However, a deauth attack is more likely to be detected since the victim will experience multiple disconnection and connection attempts.

The reason why this attack is possible is because management frames are sent in the clear. This is because these frames are used to broadcast an SSID or to initiate a connection, thus a prospective client will need to be able to read these frames even before authentication and key negotiation. Therefore, these frames cannot be encrypted.

\subsection{Captive portals}

The captive portal attack exploits a vulnerability on most iOS/OSX devices running a certain version. While the challange response phase is taking place, the radius server will send a TLV-success packet. The client device is supposed to restart the connection attempt since the challenge response was not completed. However, a vulnerable device will respond with a TLV-success packet as well. The client device will then check for the existence of a captive portal and load it in the browser. A victim who is not technologically savvy or familiar with the AP may then enter his username and password which will be transmitted in the clear to the server. 

However, this attack is much more intrusive and easily detectable since a user who is familiar with the GWireless network will have never seen the captive portal before and will find it suspicious. That being said, this attack\cite{8} was carried out at DEFCON 21, a conference for hackers and security professionals, managed to trick a number of them into revealing their credentials.

%------------------------------------------------

\section{Ethical considerations}

I did not obtain permission from any of the victims to carry out this attack. In most cases, I did not even know who the victim was, thus it is difficult to obtain permission from any of them. I limited myself to carrying out only the credentials collection portion of the attack as I did not want to possess the cleartext password for any other user. For the same reason, I did not perform the captive portal attack as it would reveal the cleartext password. I also decided against performing the deauthentication attack since it is will cause a loss in network connectivity and disruption to all users in the vicinity.

%------------------------------------------------

\section{Recommendations}

To prevent an Evil Twin attack, GWU can upload the certificate used by their radius server so that it can be downloaded and used by all users who are concerned over their privacy. By doing so, they will not inconvenience the general public who can choose not to supply a certificate but allow concerned users to do so. Users who have configured the certificate will not be vulnerable to the evil twin attack.

\end{multicols}
\includegraphics[width=\textwidth]{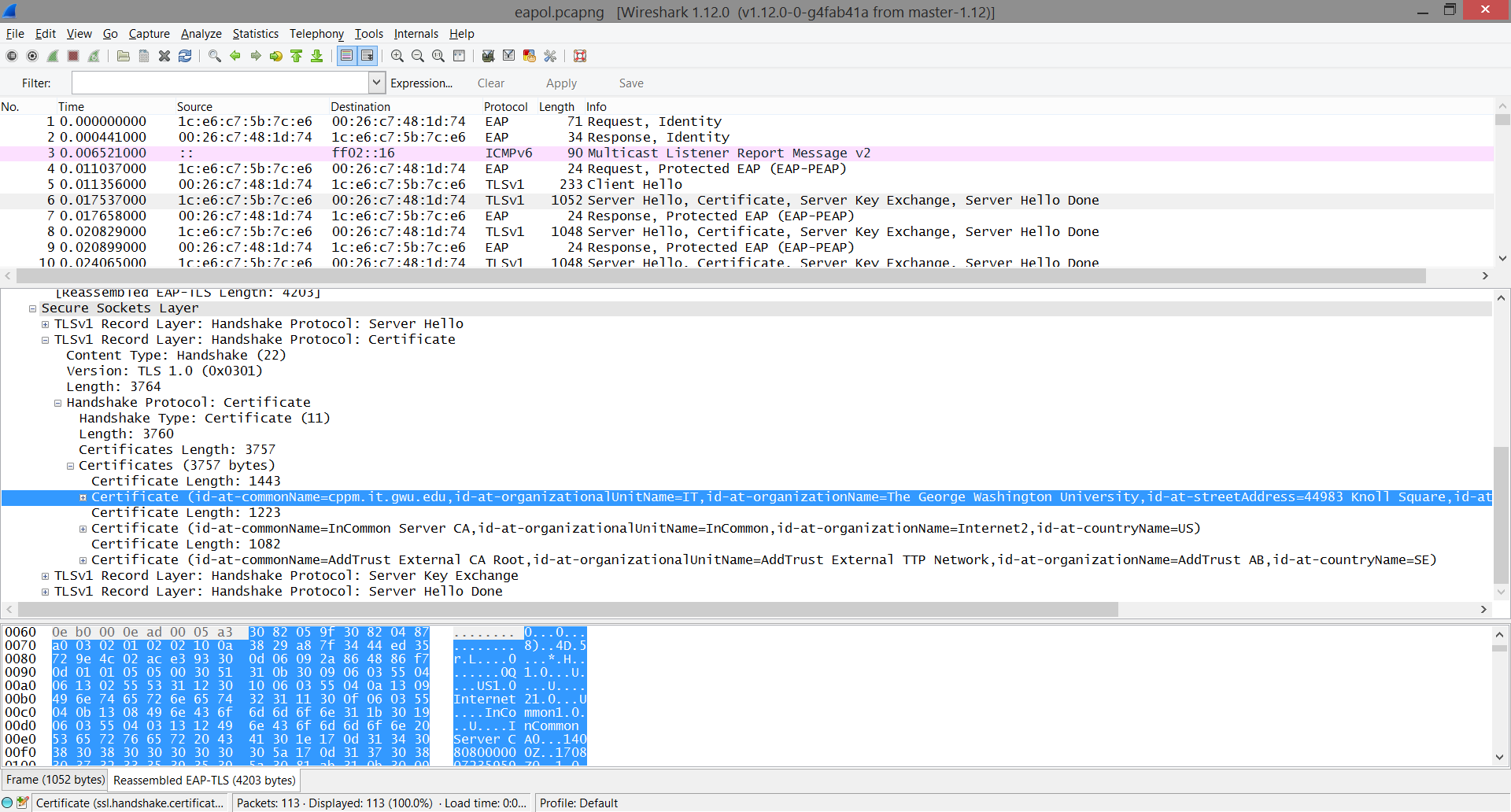}
\begin{multicols}{2}

As depicted in the screenshot above, I obtained the certificate chain by doing a packet capture while connecting to a real access point. After configuring my devices with the certificates, I found that they were not longer trying to authenticate with the rogue AP, thus proving that the recommendation would work. The capture file and certificates will be made available together with this report. 

Even though EAP-TLS would be ideal, I am against recommending it. This is because a user certificate would have to be generated for every user and the less tech savvy would have problems configuring their devices to connect to GWireless using the certificates. As a result, there would be a huge problem with certificate management and technical support. EAP-TLS would be better suited to a smaller organisation which has a greater need for security.

In the long term, large organizations like GWU would have to move away from MS-CHAPv2 towards a protocol that is more resistant against a brute force attack. MS-CHAPv2 uses DES to encrypt the challenges and MD4 to hash the passwords, both of which are algorithms of yesterday. We would need a protocol that incorporates more recent algorithms like AES and SHA-2 to stay ahead of the game.

%------------------------------------------------

\section{Conclusion}

This report has shown that it is feasible to stealthily recover user credentials from the GWireless network using off-the-shelf equipment. Although I did not demonstrate cracking the hashes to retrieve plaintext credentials, I have outlined the steps and provided data from other research that shows its feasibility. The report has also briefly touched on various variations that are slightly more intrusive but likely to give a better result. Lastly, I have also proposed a recommendation that would optionally improve security for privileged users while still not inconveniencing the general populace. 

%------------------------------------------------
\end{multicols}

%----------------------------------------------------------------------------------------
%	REFERENCE LIST
%----------------------------------------------------------------------------------------
\pagebreak

%----------------------------------------------------------------------------------------


\begin{thebibliography}{99} % Bibliography - this is intentionally simple in this template

\bibitem[1]{1}
Gartner (2015).\\
\newblock Gartner Predicts by 2017, Half of Employers will Require Employees to Supply Their Own Device for Work Purposes\\
\newblock http://www.gartner.com/newsroom/id/2466615\\
\newblock {\em Retrieved 10 April 2015}.

\bibitem[2]{2}
Yue Ma, and Xiuying Cao.(2003).\\
\newblock How to use EAP-TLS authentication in PWLAN environment. In Neural Networks and Signal Processing, 2003. Proceedings of the 2003 International Conference on (Vol. 2, pp. 1677-1680). IEEE.

\bibitem[3]{3}
GWU Division of Information Technology \\
\newblock Network \& Internet Access | Division of IT | The George Washington University\\
\newblock it.gwu.edu/network-internet-access\\
\newblock {\em Retrieved 10 April 2015}.

\bibitem[4]{4}
The FreeRADIUS Server Project and Contributors \\
\newblock FreeRADIUS: The world's most popular RADIUS Server\\
\newblock http://freeradius.org/\\
\newblock {\em Retrieved 10 April 2015}.

\bibitem[5]{5}
Cassola, A., Robertson, W. K., Kirda, E., \& Noubir, G. (2013, February).\\
\newblock  A Practical, Targeted, and Stealthy Attack Against WPA Enterprise Authentication. In NDSS.

\bibitem[6]{6}
Pico Computing, Inc.\\
\newblock Pico Computing: Solutions for FPGA and Embedded HPC Applications\\
\newblock http://picocomputing.com/\\
\newblock {\em Retrieved 10 April 2015}.

\bibitem[7]{7}
Aircrack-ng\\
\newblock Aircrack-ng\\
\newblock http://www.aircrack-ng.org/\\
\newblock {\em Retrieved 10 April 2015}.

\bibitem[8]{8}
Joel Voss. (2013)\\
\newblock PEAP at DEF CON 21 | Leviathan Security Group\\
\newblock https://www.leviathansecurity.com/blog/peap-at-def-con-21/\\
\newblock {\em Retrieved 10 April 2015}.

\end{thebibliography}
\end{document}